\documentclass[10pt,conference]{IEEEtran}

\usepackage{amsmath,amssymb,latexsym}
\usepackage{stmaryrd}
\usepackage{amsfonts}
\usepackage{times,lastpage,bm,xspace}
\usepackage{bibunits}
\usepackage[pdftex]{graphicx}

\usepackage{theorem}
\newtheorem{thm}{Theorem}
\usepackage[usenames]{color}

\definecolor{plum}  {rgb}{.4,0,.4}

 \newcommand{\xmath}[1]{{\ensuremath{#1}\xspace}}

 \def \Poisson{{\xmath{{\rm Poisson}}}}
 \def\tf{f^*}

\def \ttheta{\theta^*}
\def\tA{\widetilde{A}}

\def\pen{\mathop{\rm pen}}
\def\deq{\triangleq}
\def\thetak{\theta^{(k)}}

\def\hf{\widehat{f}}

\def\KL{\mathop{\rm KL}}

\def\jth{{\xmath{j^{{\rm th}}}}\xspace}

 \def\argmin{\mathop{\rm arg\,min}}
 \newcommand{\reals}{{\xmath{\mathbb{R}}}}
 
 \newcommand{\expect}{{\xmath{\mathbb{E}}}}
 \newcommand{\prob}{{\xmath{\mathbb{P}}}}

\newcommand{\squishlist}{
   \begin{list}{$\bullet$}
    { \setlength{\itemsep}{0pt}      \setlength{\parsep}{0pt}
      \setlength{\topsep}{0pt}       \setlength{\partopsep}{0pt}
      \setlength{\leftmargin}{1.5em} \setlength{\labelwidth}{1em}
      \setlength{\labelsep}{0.5em} } }

\newcommand{\squishlisttwo}{
   \begin{list}{$\bullet$}
    { \setlength{\itemsep}{0pt}    \setlength{\parsep}{0pt}
      \setlength{\topsep}{0pt}     \setlength{\partopsep}{0pt}
      \setlength{\leftmargin}{2em} \setlength{\labelwidth}{1.5em}
      \setlength{\labelsep}{0.5em} } }

\newcommand{\squishend}{
    \end{list}  }

\def\cF{{\cal F}}
\def\cG{{\cal G}}
\def\cC{{\cal C}}
\def\cS{{\cal S}}

\def\ave#1#2{\langle #1, #2 \rangle}
\def\1{{\bf 1}}

\begin{document}

\title{Performance Bounds on Compressed Sensing\\
with Poisson Noise}

\author{\authorblockN{Rebecca M. Willett}
\authorblockA{Electrical and Computer Engineering\\
Duke University\\
Durham, NC 27708, USA\\
Email: willett@duke.edu}
\and
\authorblockN{Maxim Raginsky}
\authorblockA{Electrical and Computer Engineering\\
Duke University\\
Durham, NC 27708, USA\\
Email: m.raginsky@duke.edu}}
\maketitle

\begin{abstract}
  This paper describes performance bounds for compressed sensing in
  the presence of Poisson noise when the underlying signal, a vector
  of Poisson intensities, is sparse or compressible (admits a sparse
  approximation). The signal-independent and bounded noise models used
  in the literature to analyze the performance of compressed sensing
  do not accurately model the effects of Poisson noise. However,
  Poisson noise is an appropriate noise model for a variety of
  applications, including low-light imaging, where sensing hardware is
  large or expensive, and limiting the number of measurements
  collected is important. In this paper, we describe how a feasible
  positivity-preserving sensing matrix can be constructed, and then
  analyze the performance of a compressed sensing reconstruction approach for
  Poisson data that minimizes an objective function consisting of a
  negative Poisson log likelihood term and a penalty term which could
  be used as a measure of signal sparsity.
\end{abstract}

\section{Introduction}
\label{sec:intro}

The basic idea of compressed sensing is that, when the signal of
interest is very sparse (i.e.,\ zero-valued at most locations) or
highly compressible in some basis, relatively few ``incoherent''
observations are sufficient to reconstruct the most significant
non-zero signal components \cite{CS:candes2,CS:donoho}. Despite the
promise of this theory for many applications, very little is known
about its applicability to photon-limited imaging systems, where
high-quality photomultiplier tubes (PMTs) are expensive and physically
large, limiting the number of observations that can reasonably be
collected by an imaging system.  Limited photon counts arise in a wide
variety of applications, including infrared imaging, nuclear medicine,
astronomy and night vision, where the number of photons collected by
the detector elements is very small relative to the number of pixels,
voxels, or other quantities to be estimated. Robust reconstruction
methods can potentially lead to many novel imaging systems designed to
make the best possible use of the small number of photons collected
while reducing the size and cost of the detector array.

However, the signal-independent and bounded noise models which have
been considered in the literature (cf. \cite{CS:noiseEC,CS:noiseRN})
are not easily adapted to the Poisson noise models used in
photon-limited imaging.  The Poisson model is often used to model
images acquired by photon-counting devices \cite{SHW93}. Under the
Poisson assumption, we can write our observation model as
\begin{equation}
y \sim \Poisson(A\tf), \label{eq:Poisson2}
\end{equation}
where $\tf \in \reals^{m}$ is the signal or image of interest, $A \in
\reals^{N \times m}$ linearly projects the scene onto an
$N$-dimensional space of observations, and $y \in \{0,1,2,\ldots\}^N$ is a
length-$N$ vector of observed Poisson counts. Specifically, under the model in (\ref{eq:Poisson2}), the likelihood of observing a particular vector of counts $y$ is given by
$$
p(y|A\tf) = \prod^N_{j=1} \frac{(A\tf)_j^{y_j}}{y_j!} e^{-(A\tf)_j},
$$
where $(A\tf)_j$ is the \jth component of $A\tf$.

The majority of the compressed sensing literature assumes that there
exists a ``sparsifying'' reference basis $W$, so that $\ttheta \deq
W^T \tf$ is sparse or lies in a weak-$\ell_p$ space.  When the matrix
product $AW$ obeys the so-called {\em restricted isometry property}
(RIP) \cite{RIP,JustRelax} or some related criterion, and when the
noise is bounded or Gaussian, then $\ttheta$ can be accurately
estimated from $y$ by solving the following $\ell_2-\ell_1$
optimization problem (or some variant):
\begin{equation}
\widehat{\theta} 
= 
\argmin_{\theta}  \|y-AW\theta\|_2^2 + \tau \|\theta\|_1,
\label{eq:nlp2}
\end{equation}
where $\tau > 0$ is a regularization parameter
\cite{CS:donoho,LASSO,JustRelax}. 

However, the $\ell_2$ data-fitting term, $\|y-AW\theta\|_2^2$, is
problematic in the presence of Poisson noise.  Because the variance of
the noisy observations is proportional to the signal intensity,
$\ell_2$ data-fitting terms can lead to significant overfitting in
high-intensity regions and oversmoothing in low-intensity regions.
Furthermore, photon-limited imaging systems implicitly place hard
constraints on the nature of the measurements that can be collected,
such as non-negativity, which are not considered in much of the
existing compressed sensing literature (recent papers of Dai and
Milenkovic \cite{DaiMil09} and of Khajehnejad et al. \cite{KDXH09} are
notable exceptions).

In this paper, we propose estimating $\tf$ from $y$ using a
regularized Poisson log-likelihood objective function as an
alternative to (\ref{eq:nlp2}), and we present risk bounds for
recovery of a compressible signal from Poisson
observations. Specifically, in the Poisson noise setting we maximize
the log-likelihood while minimizing a penalty function that, for
instance, could measure the sparsity of $\theta=W^Tf$:
\begin{equation}
\begin{array}{rll}
\displaystyle
	\hf \ = \ & \displaystyle \argmin_{f}  & \displaystyle
	 \sum_{j=1}^{N} \left (-y_j \log (Af)_j \right ) + \tau \pen( f ) \\
	& \textrm{subject to} &  Af \succeq 0, \;        f
        \succeq 0, \;  \sum_{i=1}^m f_i = I
\end{array} \label{eq:CSP2} 
\end{equation}
where $\pen(\cdot)$ is a penalty function that will be detailed later, $I$ is the total intensity of the unknown $\tf$ (assumed known), and the standard notation $v \succeq 0$ means that the components of
$v$ are nonnegative.  The constraints reflect the nonnegativity of
both the observed intensity and the underlying image and the known
total intensity of the underlying image.

\section{Problem formulation}
\label{sec:problem}

We have a signal or image $\tf$ of length $m$ that we wish to
estimate using a detector array of length $N \ll m$. We assume that $\tf \succeq 0$. We will bound the accuracy
with which we can estimate $\tf/I$, where $I \deq
\sum_{i=1}^m \tf_i$; in other words, we focus on accurately estimating
the {\em shape} of $\tf$ independent of any scaling factor
proportional to the total intensity of the scene. We assume that the total intensity $I$ is known, and our candidate estimators will also be constrained to have total intensity $I$. The quality of a candidate estimator $f$ will be measured in terms of the {\em risk}
$$
R(\tf,f) \deq \left\| \frac{\tf}{I} - \frac{f}{I} \right \|^2_2.
$$

We construct our sensing matrix $A$ as follows. Let $Z \in \{-1,+1\}^{N \times m}$ be a matrix whose entries
$Z_{i,j}$ are independent Rademacher random variables, i.e., $\prob
\left[Z_{i,j} = -1 \right] = \prob \left[Z_{i,j} = +1 \right] =
1/2$ independently of all other $Z_{i',j'}$. Let $\tA =
(1/N)Z$. Most compressed sensing approaches would proceed by assuming that we
make (potentially noisy) observations of the product $\tA \tf$, but
elements of $\tA \tf$ could be negative and thus not physically
realizable in photon-counting systems. However, we can use $\tA$ to
generate a positivity-preserving sensing matrix $A$ as follows. Let $\1_{r \times s}$ denote the $r \times s$ matrix all of whose entries are equal to 1. Then we let
$$
A \deq \tA + (1/N) \1_{N \times m}.
$$
Note that $A \in \{0,2/N\}^{N \times m}$ and, as a consequence, $A$ indeed preserves positivity: for any $f \in \reals^m_+$, $Af \succeq 0$.

We make Poisson observations of $A\tf$, $y \sim \Poisson(A\tf)$, and our goal is to estimate $\tf \in \reals^m_+$ from $y \in
\{0,1,2,\ldots\}^N$. To this end, we propose solving
the following optimization problem:
\begin{equation}
\hf \deq \argmin_{f \in \Gamma} \Big[ -\log p(y|Af) +
  2\pen(f) \Big],
\label{eq:opt}
\end{equation}
where $\pen(f)$ is a penalty term.
We assume that $\Gamma \equiv \Gamma(m,I)$ is a countable set of feasible estimators $f \in \reals^m_+$
satisfying $\sum^m_{i=1} f_i = I$, and that the penalty function satisfies the Kraft inequality:
\begin{equation}
\sum_{f \in \Gamma} e^{-\pen(f)} \leq 1. \label{eq:kraft}
\end{equation}
Note that, by construction of $A$, $f \in \Gamma$ implies that $Af \succeq 0$. Furthermore, while the penalty term
may be chosen to be smaller for sparser solutions $\theta = W^T f$, where $W$ is an orthogonal matrix that represents $f$ in its ``sparsifying" basis, our main
result only assumes that (\ref{eq:kraft}) is satisfied. We can think
of (\ref{eq:opt}) as a discretized-feasibility version of
(\ref{eq:CSP2}), where we optimize over a countable set of feasible
vectors that grows in a controlled way with signal length $m$.

\section{Properties of the sensing matrix $A$}
\label{sec:A_prop}

Our main result, stated and proved in the next section, makes use of the several properties of the sensing matrix $A$ (and $\tA$). The most important of these properties is that, with high probability, $\tA$ acts near-isometrically on certain subsets of $\reals^m$. The usual formulation of this phenomenon is known in the compressed sensing literature as the {\em restricted isometry property} (RIP) \cite{RIP,JustRelax}, where the subset of interest consists of all vectors with a given sparsity. In fact, the RIP is a special case of a much broader circle of results concerning the behavior of random matrices whose entries are drawn from a subgaussian isotropic ensemble \cite{MenPajTom07}. The Rademacher ensemble is an instance of this, and the following two theorems can be extracted from the results of \cite{MenPajTom07}:

\begin{thm}\label{thm:subgauss_1} There exist absolute constants $c_1,c_2 > 0$, such that, with probability at least $1-e^{-c_1 N}$,
$$
\| u - v \|_2 \le \sqrt{2} N \| \tA ( u - v) \|_2 + c_2 \sqrt{\frac{\log (c_2m/N)}{N}}
$$
for all $u,v \in \reals^m$ such that $\| u \|_1 = \| v \|_1 = 1$.
\end{thm}

\begin{thm}\label{thm:subgauss_2} There exist absolute constants $c_3,c_4 > 0$, such that the following holds. Let $\cS$ be a finite subset of the unit sphere in $\reals^m$. Then, with probability at least $1-e^{-c_3 N}$,
$$
1/2 \le N \| \tA s \|^2_2 \le 3/2, \qquad \forall s \in \cS
$$
provided $N \ge c_4 \log_2 |\cS|$.
\end{thm}

We will also rely on the following properties of $A$ and $\tA$:
\begin{itemize}
\item With probability at least $1-N2^{-m}$, every row of $Z$ has at least one positive entry. Let $f \in \reals^m$ be an arbitrary vector of intensities satisfying $f \succeq (cI)\1_{m \times 1}$ for some $c > 0$. Then
\begin{equation}
A f \succeq (2cI/N)\1_{N \times 1}.
\label{eq:pos_intens}
\end{equation}
\item With probability at least $1-2m e^{-N/8}$,
\begin{equation}
\left| \sum^N_{i=1} \tA_{i,j} \right| \le 1/4, \qquad \forall j \in \{1,\ldots,m\}
\label{eq:col_sums}
\end{equation}
(this is a simple consequence of the Chernoff bound and the union bound).
\item If the event (\ref{eq:col_sums}) holds, then
\begin{equation}
(3/4)I \le \sum^N_{i=1} \sum^m_{j=1} A_{i,j} f_j \le (5/4)I, \qquad \forall f \in \reals^m_+
\label{eq:det_intensity}
\end{equation}
\end{itemize}

\section{An oracle inequality for the expected risk}
\label{sec:oracle}

We now state and prove our main result, which gives an upper bound on
the expected risk $\expect R(\tf,\hf)$ that holds for any target
signal $\tf \succeq 0$ satisfying the normalization constraint
$\sum^m_{i=1} \tf_i = I$, without assuming anything about the
sparsity properties of $\tf$. Conceptually, our bound is an {\em
  oracle inequality}, which states that the expected risk of our
estimator is within a constant factor of the best regularized risk
attainable by estimators in $\Gamma$ with full knowledge of the
underlying signal $\tf$. More precisely, for each $f \in \Gamma$ define
$$
R^*(\tf,f) \deq \left\| \frac{\tf}{I} - \frac{f}{I} \right\|^2_2 + \frac{2\pen(f)}{I},
$$
and for every $\Gamma' \subseteq \Gamma$ let $R^*(\tf,\Gamma') \deq \min_{f \in \Gamma'} R^*(\tf,f)$. Note that $R^*(\tf,\Gamma')$ is the best penalized risk that can be attained over $\Gamma'$ by an oracle that has full knowledge of $\tf$. We then have the following:

\begin{thm}\label{thm:main} Suppose that the feasible set $\Gamma$
  also satisfies the condition
\begin{equation}
f \succeq (cI) \1_{m \times 1}, \qquad \forall f \in \Gamma 
\label{eq:positivity}
\end{equation}
for some $0 < c < 1$. Let $\cG_N$ be the collection of all subsets $\Gamma' \subseteq \Gamma$, such that $|\Gamma'| \le 2^{N/c_4}$. Then the following holds with probability at least $1-me^{-KN}$ for some positive $K=K(c_1,c_3)$ (with respect to the realization of $\tA$):
\begin{equation}
\expect R(\tf,\hf) \le C_N \min_{\Gamma' \in \cG_N} R^*(\tf,\Gamma') + \frac{2c^2_2 \log (c_2 m/N)}{N},
\label{eq:thm}
\end{equation}
where $C_N = \max(20,15/c)N$, and the expectation is taken with respect to $y \sim \Poisson(A\tf)$.
\end{thm}

\noindent{{\bf Remark 1.} A positivity condition similar to
  (\ref{eq:positivity}) is natural in the context of estimating
  vectors with nonnegative entries from count data. In particular, it
  excludes the possibility of assigning zero intensity to an input of
  a detector when at least one photon has been counted
  \cite{Csi91}. However, as will be clear from the proof below, condition (\ref{eq:positivity}) can be replaced with a more general (weaker) condition
  $$
  Af \succeq (c' I/N) \1_{N \times 1}, \qquad \forall f \in \Gamma
  $$
  for some $c' > 0$, which is more appropriate when the signal $\tf$ is sparse in the canonical basis, because then it is in fact desirable to allow candidate estimators with zero components.\\

\begin{proof}
With high probability, the following chain of estimates holds:
\begin{eqnarray*}
\lefteqn{\frac{1}{I^2} \| \tf - \hf \|^2_2} \\
&\le& \frac{4N}{I^2} \| \tA (\tf - \hf) \|^2_2 + \frac{2c^2_2\log (c_2 m/N)}{N}  \\
&=& \frac{4N}{I^2} \| A (\tf - \hf) \|^2_2 + \frac{2c^2_2\log (c_2 m/N)}{N}   \\
&\le& \frac{4N}{I^2} \| A (\tf - \hf) \|^2_1 + \frac{2c^2_2\log (c_2 m/N)}{N},
\end{eqnarray*}
where the first inequality is a consequence of Theorem~\ref{thm:subgauss_1}, and the remaining steps follow from definitions and from standard inequalities for $\ell_p$ norms. Moreover, with high probability,
\begin{eqnarray*}
\lefteqn{\| A (\tf - \hf) \|^2_1 } \\
&=& \left(\sum^N_{i=1} \left| (A\tf)^{1/2}_i - (A\hf)^{1/2}_i \right| \cdot \left|(A\tf)^{1/2}_i +  (A\hf)^{1/2}_i \right|   \right)^2 \\
&\le& \sum^N_{i,j=1} \left|(A\tf)^{1/2}_i - (A\hf)^{1/2}_i \right|^2 \left|(A\tf)^{1/2}_j +  (A\hf)^{1/2}_j \right|^2 \\
&\le& 5 I \sum^N_{i=1} \left|(A\tf)^{1/2}_i - (A\hf)^{1/2}_i \right|^2,
\end{eqnarray*}
where the first inequality is due to Cauchy--Schwarz, and the second inequality is a consequence of (\ref{eq:det_intensity}) and the inequality between the arithmetic mean and the geometric mean. It is a matter of straightforward algebra to show that
\begin{eqnarray*}
\lefteqn{\sum^N_{i=1} \left|(A\tf)^{1/2}_i - (A\hf)^{1/2}_i \right|^2} \\
&=& -2 \log \sum^N_{i=1} \exp \left(-\frac{1}{2} \left[ (A\tf)^{1/2}_i - (A\hf)^{1/2}_i \right]\right)^2 \\
&=& 2 \log \left(\int \sqrt{ p(y|A\tf) p(y|A\hf)} d\nu(y)\right)^{-1},
\end{eqnarray*}
where $\nu$ is the counting measure on $\{0,1,2,\ldots\}^N$. Now,  the same techniques as in Li and Barron \cite{libarron} (see also the proof of Theorem~7 in \cite{kola2}) can be used to show that 
\begin{eqnarray}
\lefteqn{2 \expect \log \left(\int \sqrt{ p(y|A\tf) p(y|A\hf)} d\nu(y)\right)^{-1} } \nonumber \\
&\le& \min_{f \in \Gamma} \left[ \KL\Big(p(\cdot|A\tf) \Big\| p(\cdot|Af)\Big) + 2\pen(f) \right],
\label{eq:li_barron}
\end{eqnarray}
where $\KL(\cdot \| \cdot)$ is the {\em Kullback--Leibler (KL) divergence}, which for the Poisson likelihoods has the form
\begin{eqnarray*}
\lefteqn{\KL\Big(p(\cdot|A\tf) \Big\| p(\cdot|Af)\Big)} \\
& =& \sum^N_{i=1} \left[(A\tf)_i \log \frac{(A\tf)_i}{(Af)_i} - (A\tf)_i + (Af)_i\right].
\end{eqnarray*}
Using the inequality $\log t \le t -1$ together with (\ref{eq:positivity}) and (\ref{eq:pos_intens}), we can bound the KL divergence as
\begin{eqnarray*}
\lefteqn{\sum^N_{i=1} \left[(A\tf)_i \log \frac{(A\tf)_i}{(Af)_i} - (A\tf)_i + (Af)_i\right]} \\
&\le& \sum^N_{i=1} \left[(A\tf)_i \left( \frac{(A\tf)_i}{(Af)_i} - 1\right) - (A\tf)_i + (Af)_i\right] \\
&=& \sum^N_{i=1} \frac{1}{(Af)_i} \left[ (Af)^2_i - 2(Af)_i (A\tf)_i + (A\tf)^2_i\right] \\
&\le& \frac{N}{2cI} \| A(\tf - f) \|^2_2 \\
&=& \frac{N}{2cI} \| \tA (\tf - f ) \|^2_2.
\end{eqnarray*}
Now, choose any $\Gamma^* \in \cG_N$, such that
$$
R^*(\tf,\Gamma^*) = \min_{\Gamma' \in \cG_N} R^*(\tf,\Gamma').
$$
Then, applying Theorem~\ref{thm:subgauss_2} to the set $\left\{ \frac{\tf - f}{\| \tf - f \|_2} : f \in \Gamma^*\right\}$, we have, with high probability, that
$$
N \| \tA(\tf - f) \|^2_2 \le (3/2) \| \tf - f \|^2_2, \qquad \forall f \in \Gamma^*.
$$
Combining everything, we get the bound $\expect R(\tf,\hf) \le$
\begin{eqnarray*}
&& \max\left(20,\frac{15}{c}\right) N \min_{f \in \Gamma^*} \left[\left\| \frac{\tf}{I} - \frac{f}{I} \right\|^2_2 + \frac{2\pen(f)}{I} \right] \\
&& \qquad \qquad + \frac{2c^2_2 \log (c_2 m/N)}{N}
\end{eqnarray*}
which holds with high probability w.r.t.\ the realization of $\tA$. Let $C_N = \max(20,15/c)N$. The theorem is proved. \end{proof}

\section{Risk bounds for compressible signals}
\label{sec:rates}

We now show how the bound in Theorem~\ref{thm:main} can be used to analyze how the
performance of the proposed estimator when the target signal $\tf$ is compressible (i.e.,~admits a sparse approximation) in some reference orthonormal basis.

Following \cite{CS:candes2}, we assume that there exists an orthonormal basis $\Phi = \{\phi_1,\ldots,\phi_m\}$ of $\reals^m$, such that $\tf$ is {\em compressible} in $\Phi$ in the following sense. Let $W$ be the orthogonal matrix with columns $\phi_1,\ldots,\phi_m$. Then the vector $\ttheta$ of the coefficients $\ttheta_j = \ave{\tf}{\phi_j}$ of $\tf$ in $\Phi$ is related to $\tf$ via $\tf = W\ttheta$. Let $\ttheta_{(1)},\ldots,\ttheta_{(m)}$ be the entries of $\ttheta$ arranged in the order of decreasing magnitude: $|\ttheta_{(1)}| \ge |\ttheta_{(2)}| \ge \ldots \ge |\ttheta_{(m)}|$. We assume that there exist some $0 < q < \infty$ and $\rho > 0$, such that for each $1 \le j \le m$
\begin{equation}
|\ttheta_{(j)}| \le \rho I j^{-1/q}.
\label{eq:weak_lp}
\end{equation}
Note that for every $1 \le j \le m$ we have
$$
|\ttheta_{(j)}| \le \| \ttheta \|_2 = \| \tf \|_2 \le \| \tf \|_1 = I,
$$
so we can take $\rho$ to be a constant independent of $I$ or $m$. Any $\ttheta$ satisfying (\ref{eq:weak_lp}) is said to belong to the {\em weak-$\ell_q$ ball of radius $\rho I$}. The weak-$\ell_q$ condition (\ref{eq:weak_lp}) translates into the following approximation estimate: given any $1 \le k \le m$, let $\thetak$ denote the best $k$-term approximation to $\ttheta$. Then we can show that
\begin{equation}
\left \| \frac{\ttheta}{I} - \frac{\thetak}{I} \right\|^2_2 \le C \rho^2 k^{-2\alpha}, \qquad \alpha = 1/q - 1/2
\label{eq:compressibility}
\end{equation}
for some constant $C > 0$ that depends only on $q$. We also assume that $\tf$
satisfies the condition (\ref{eq:positivity}) for some $c \in (0,1)$, a
lower bound on which is assumed known.

In order to apply Theorem~\ref{thm:main}, we will form a
suitable finite class of estimators $\Gamma$ and set a penalty function
$\pen(f)$ over this class which (a) is smaller for sparser
$\theta = W^T f$ and (b) satisfies (\ref{eq:kraft}).  The family $\Gamma$ is
constructed as follows.
\begin{enumerate}
\item Define the sets
\begin{eqnarray*}
\lefteqn{ \Theta \deq \left\{ \theta \in \reals^m : \| \theta \|_\infty
    \le I; \right. } \\
&& \left.  \mbox{ each $\theta_i$ uniformly quantized to one of $\sqrt{m}$
    levels} \right\} 
\end{eqnarray*}
and $\cF \deq \left\{ f \in \reals^m : f = W\theta, \theta \in \Theta \right\}$.
\item For each $f \in \cF$, let $\bar{f}$ denote the $\ell_2$
  projection of $f$ onto the closed convex set
  $$
  \cC \deq \left\{ g \in \reals^m: g \succeq (cI)\1_{m \times 1} \mbox{ and } \sum^m_{i=1} g_i = I \right\},
  $$
i.e.,
$$
\bar{f} \deq \argmin_{g \in \cC} \| f - g \|_2.
$$
\item Finally, let $\Gamma \deq \left\{ \bar{\theta} = W^T \bar{f} : f \in \cF \right\}$.
\end{enumerate}
Note that the projection $\bar{f}$ satisfies the {\em Pythagorean identity}
$$
\| g - f \|^2_2 \ge \| g - \bar{f} \|^2_2 + \| \bar{f} - f \|^2_2, \qquad \forall g \in \cC
$$
(see, e.g.,~Theorem~2.4.1 in \cite{CenZen97}). In particular, $\| g - f \|^2_2 \ge \| g - \bar{f} \|^2_2$, and, since $\tf \in \cC$, we have
\begin{equation}
\| \tf - \bar{f} \|^2_2 \le \| \tf - f \|^2_2, \qquad\forall f \in \cF.
\label{eq:projection}
\end{equation}

Consider the penalty
$$
\pen(f) = \log_2(m+1) +  (3/2) \|\theta\|_{0} \log_2(m) , \qquad \theta = W^T f.
$$
This corresponds to the following prefix code for $\theta \in \Theta$
(that is, we encode the elements of $\Theta$, before they are
subjected to the deterministic operation of projecting onto $\cC$):
\begin{enumerate}
\item First we encode $\| \theta \|_0$, the number of nonzero components of $\theta$, which can
  be encoded with $\log_2(m+1)$ bits.
\item For each of the $\|\theta\|_{0}$ nonzero components, we store
  its location in the $\theta$ vector; since there are $m$ possible
  locations, this takes $\log_2(m)$ bits per component.
\item Next we encode each coefficient value, quantized to one of
  $\sqrt{m}$ uniformly sized bins.
\end{enumerate}
Since this corresponds to a uniquely decodable code for $f \in \cF$ (or $\theta \in \Theta)$, we see that $\pen(f)$ satisfies
the Kraft inequality.

Now, given $\ttheta = W^T \tf$, let $\thetak$ be its best $k$-term approximation,  $\thetak_{q} \in \Theta$ the quantized version of
$\thetak$, for which we have
$$
\left\| \frac{\thetak_{q}}{I} - \frac{\thetak}{I} \right\|^2_2 \le \frac{k}{m},
$$
and $\bar{\theta}^{(k)}_q$ the  element of
$\Gamma$ obtained by projecting $f^{(k)}_q = W \thetak_q$ onto $\cC$ and then transforming back into the basis $\Phi$: $\bar{\theta}^{(k)}_q = W^T \bar{f}^{(k)}_q$. Then, using (\ref{eq:projection}) and (\ref{eq:compressibility}), we get
\begin{eqnarray*}
\| \tf  - \bar{f}^{(k)}_q \|^2_2 &\le& \| \tf - f^{(k)}_q \|^2_2 \\
  &=& \| \ttheta - \thetak_q \|^2_2 \\
&\le& 2 \| \ttheta - \thetak \|^2_2 + 2 \| \thetak - \thetak_q \|^2_2 \\
&\le& I^2 \left( 2Ck^{-2\alpha} + \frac{2k}{m} \right).
\end{eqnarray*}
Given each $1 \le k \le m$, let $\Gamma_k \subseteq \Gamma$ be the set of all $\bar{\theta} \in \Gamma$, such that the corresponding $\theta \in \Theta$ satisfies $\| \theta \|_0 \le k$. Then $|\Gamma_k| = {m \choose k} m^{k/2}$, so that $\log_2 |\Gamma_k| \le 2k \log_2 m$, and therefore $\Gamma_k \in \cG_N$ whenever $k \le k_*(N)$, where $k_*(N) \deq N/(2c_4 \log_2 m)$. Then the first term on the right-hand side of (\ref{eq:thm}) can be bounded by
\begin{eqnarray*}
\lefteqn{C_N \min_{1 \le k \le k_*(N)} R^*(\tf,\Gamma_k)} \\
&\le& O(N) \min_{1 \le k \le k_*(N)} \left[ \left\| \frac{\ttheta}{I} - \frac{\bar{\theta}^{(k)}_q}{I} \right\|^2_2 + \frac{2 \pen(f^{(k)}_q)}{I}\right] \\
&\le& O(N) \min_{1 \le k \le k_*(N)} \left[ k^{-2\alpha} + \frac{k}{m} + \frac{k \log_2 m}{I}\right],
\end{eqnarray*}
where the constant obscured by the $O(\cdot)$ notation depends only on $C$ and $c$.
We can now consider two cases:

\noindent 1) $I \le m\log m$, i.e., the penalty term dominates the quantization error, then we get the risk bound $\expect R(\tf,\hf) \le$
$$
O(N) \min_{1 \le k \le k_*(N)} \left[k^{-2\alpha}  + \frac{2k \log_2 m}{I}\right]  + \frac{2c^2_2 \log(c_2m/N)}{N}.
$$
If $k_*(N) \ge (\alpha I/\log_2 m)^{1/(2\alpha+1)}$, then we can further obtain
$$
\expect R(\tf,\hf) \le O(N) \left( \frac{I}{\log m} \right)^{-\frac{2\alpha}{2\alpha + 1}} + \frac{2c^2_2 \log(c_2m/N)}{N}.
$$
If $k_*(N) < (\alpha I/\log_2 m)^{1/(2\alpha +1)}$, there are not enough measurements, and the estimator saturates, although its risk can be controlled.

\noindent 2) $I > m \log m$, i.e., the quantization error dominates the penalty term. Then we obtain $\expect R^*(\tf,\hf) \le$
$$
O(N) \min_{1 \le k \le k_*(N)} \left[k^{-2\alpha} + \frac{2k}{m}\right] + \frac{2c^2_2 \log (c_2 m/N)}{N}.
$$
If $k_*(N) \ge (\alpha m)^{1/(2\alpha+1)}$, then we can further get
$$
\expect R(\tf,\hf) \le O(N) m^{-\frac{2\alpha}{2\alpha + 1}} + \frac{2c^2_2 \log(c_2 m/N)}{N},
$$
Again, if $k_*(N) < (\alpha m)^{1/(2\alpha+1)}$, there are not enough measurements, and the estimator saturates. 

Note that, when $I \asymp m$ and $N \asymp m^{1/p}$ for some $p > 1 + 1/2\alpha$, we get (up to log factors) the rates
$$
\expect R(\tf,\hf) = O \left(m^{-\beta}\right),
$$
where $\beta = \frac{2\alpha - (2\alpha + 1)/p}{2\alpha + 1} > 0$.

\section{Conclusion}
\label{sec:conclude}

We have derived upper bounds on the compressed sensing estimation
error under Poisson noise for sparse or compressible signals. We
specifically prove error decay rates for the case where the penalty
term is proportional to the $\ell_0$-norm of the solution; this form
of penalty has been used effectively in practice with a
computationally efficient Expectation-Maximization algorithm
(cf.~\cite{gehm:ddis}), but was lacking the theoretical support
provided by this paper. Furthermore, the main theoretical result of
this paper holds for any penalization scheme satisfying the Kraft
inequality, and hence can be used to assess the performance of a
variety of potential reconstruction strategies besides
sparsity-promoting reconstructions.

One significant aspect of the bounds derived in this paper is that
they grow with $N$, the size of the measurement array, which is a
major departure from similar bounds in the Gaussian or bounded-noise
settings. It does not appear that this is a simple artifact of our
analysis. Rather, this behavior can be intuitively understood to
reflect that elements of $y$ will all have similar values at low light levels,
making it very difficult to infer the relatively small variations in
$\tA\tf$.  Hence, Poisson compressed sensing using shifted Rademacher
sensing matrices is fundamentally difficult when the data are very
noisy. It may be possible to address these limitations through
alternative constructions of sensing matrices which introduce more
variation in the signal $A\tf$.

\section*{Acknowledgment}

This work was supported by NSF CAREER Award No. CCF-06-43947 and DARPA
Grant No. HR0011-07-1-003. The authors would also like to thank Robert
Calderbank, Emmanuel Cand\`es, Zachary Harmany, Sina Jafarpour, and
Roummel Marcia for many fruitful discussions.

\bibliographystyle{IEEEtran}
\bibliography{pcs.bbl}

\end{document}